\documentclass[preprint,showpacs,preprintnumbers,amsmath,amssymb]{revtex4}
\usepackage{graphicx}
\usepackage{dcolumn}
\usepackage{bm}

\font\scripti=cmmi7
\font\scriptscripti=cmmi5
\def\sib#1{\setbox0 = \hbox{\scripti #1}
  \kern-.02em\copy0\kern-\wd0
  \kern.04em\box0} 
\def\ssib#1{\setbox0 = \hbox{\scriptscripti #1}
  \kern-.02em\copy0\kern-\wd0
  \kern.04em\box0} 
\font\tenib=cmmib10 
\skewchar\tenib='177 \skewchar\tenib='177 \skewchar\tenib='177
\def\pbold#1{\setbox0 = \hbox{$ #1 $}
  \kern-.022em\copy0\kern-\wd0
  \kern.011em\copy0\kern-\wd0
  \kern.011em\copy0\kern-\wd0
  \kern.011em\copy0\kern-\wd0
  \kern.011em\box0} 

\usepackage{graphicx}
\usepackage{dcolumn}
\usepackage{bm}

\def\up{\uparrow}

\def\lesssim{\ \raise.3ex\hbox{$<$}\kern-0.8em\lower.7ex\hbox{$\sim$}\ }
\def\gesim{\ \raise.3ex\hbox{$>$}\kern-0.8em\lower.7ex\hbox{$\sim$}\ }

\begin{document}
\title{Uniform spin susceptibility and spin-gap phenomenon in the BCS-BEC crossover regime of an ultracold Fermi gas}
\author{Hiroyuki Tajima, Takashi Kashimura, Ryo Hanai, Ryota Watanabe, and Yoji Ohashi}
\affiliation{Department of Physics, Keio University, 3-14-1 Hiyoshi, Kohoku-ku, Yokohama 223-8522, Japan}
\date{\today}
\begin{abstract}
We investigate the uniform spin susceptibility $\chi_{\rm s}$ in the BCS (Bardeen-Cooper-Schrieffer)-BEC (Bose-Einstein condensation) crossover regime of an ultracold Fermi gas. Including pairing fluctuations within the framework of an extended $T$-matrix approximation, we show that $\chi_{\rm s}$ exhibits non-monotonic temperature dependence in the normal state. In particular, $\chi_{\rm s}$ is suppressed near the superfluid phase transition temperature $T_{\rm c}$ due to strong pairing fluctuations. To characterize this anomalous behavior, we introduce the spin-gap temperature $T_{\rm s}$ as the temperature at which $\chi_{\rm s}$ takes a maximum value. Determining $T_{\rm s}$ in the whole BCS-BEC crossover region, we identify the spin-gap regime in the phase diagram of a Fermi gas in terms of the temperature and the strength of a pairing interaction. We also clarify how the spin-gap phenomenon is related to the pseudogap phenomenon appearing in the single-particle density of states. Our results indicate that an ultracold Fermi gas in the BCS-BEC crossover region is a very useful system to examine the pseudogap phenomenon and the spin-gap phenomenon in a unified manner.
\end{abstract}
\maketitle

\par
\section{Introduction}
\par
Recently, the pseudogap phenomenon has attracted much attention in cold Fermi gas physics\cite{Tsuchiya,Watanabe,Chen,Hu,Mueller,Magierski,Su,Bulgac,Stewart,Gaebler,Perali,Kohl,Nascimbene2,Nascimbene,Navon}.  Although the pseudogap has been also discussed in high-$T_{\rm c}$ cuprates as a key to clarify the pairing mechanism of this system\cite{Renner,HTSC}, the complexity of this system still prevents us from clarifying the origin of this many-body phenomenon\cite{Yanase,Pines,Kampf,Chakravarty}. On the other hand, the ultracold Fermi gas system is much simpler than high-$T_{\rm c}$ cuprates. In addition, the strength of a pairing interaction in this system can be experimentally tuned by adjusting the threshold energy of a Feshbach resonance\cite{Timmermans,Giorgini,Chin,Bloch,Gurarie}. This unique property enables us to study the BCS-BEC crossover phenomenon\cite{Regal,Zwierlein,Kinast,Bartenstein}, where the character of a Fermi superfluid continuously changes from the weak-coupling BCS-type to the Bose-Einstein condensation of tightly bound molecules, with increasing the interaction strength\cite{Eagles,Leggett,Nozieres,SadeMelo,Ohashi,Perali2}. Since the intermediate coupling regime is dominated by strong-pairing fluctuations, this so-called BCS-BEC crossover region is expected to be useful for the assessment of the preformed pair scenario which has been discussed as a possible mechanism of the pseudogap phenomenon in high-$T_{\rm c}$ cuprates\cite{Yanase}.
\par
However, in contrast to our expectation, the pseudogap problem in cold Fermi gas physics is still in debate. While the recent photoemission-type experiments on $^{40}$K Fermi gases\cite{Stewart,Gaebler,Perali,Kohl} agree with the pseudogap scenario\cite{Tsuchiya,Watanabe,Chen,Hu,Mueller,Magierski,Su,Bulgac}, a local pressure experiment\cite{Nascimbene2}, as well as an experiment on the spin polarization rate\cite{Nascimbene}, on $^6$Li Fermi gases support the Fermi liquid theory. Since the latter theory is characterized by the existence of stable Fermi quasiparticles with long lifetime $\tau$\cite{AGD}, it is incompatible with a pseudogapped Fermi gas, where the formation of preformed pairs leads to short quasiparticle lifetime $\tau$. Thus, it is a crucial problem whether an ultracold Fermi gas is a Fermi liquid or a pseudogapped Fermi gas. 
\par
The pseudogap is a dip structure appearing in the single-particle density of states $\rho(\omega)$ around $\omega=0$ above the superfluid phase transition temperature $T_{\rm c}$\cite{Tsuchiya,Watanabe,Chen,Hu,Mueller,Magierski,Su,Bulgac}. Thus, direct observation of $\rho(\omega)$ would be the most effective approach to resolve the pseudogap problem in cold Fermi gas physics. In high-$T_{\rm c}$ cuprates, such a dip structure has been really observed by using the scanning tunneling spectroscopy (STS)\cite{Renner}. However, such a powerful technique does not exist in cold Fermi gas physics, which makes this problem more difficult.
\par
In this paper, as a useful physical quantity to resolve the pseudogap problem in cold Fermi gas physics, we pick up the uniform spin susceptibility $\chi_{\rm s}$ in the BCS-BEC crossover region above $T_{\rm c}$. In contrast to the density of states $\rho(\omega)$, $\chi_{\rm s}$ is observable in this gas system\cite{Sanner,Sommer,Lee}. In the case of a free Fermi gas, $\chi_{\rm s}$ is known to be proportional to $\rho(\omega=0)$ far below the Fermi temperature $T_{\rm F}$\cite{Kubo}. Thus, when this property still holds in the presence of a pairing interaction, one may detect the pseudogap in (unobservable) $\rho(\omega\simeq 0)$ through the anomaly in $\chi_{\rm s}$. We note that the suppression of $\chi_{\rm s}$ near $T_{\rm c}$ is known as the spin-gap phenomenon in the underdoped regime of high-$T_{\rm c}$ cuprates\cite{Yoshinari}, although the origin of this anomalous phenomenon, as well as its relation to the pseudogap, are still unclear. Since an ultracold Fermi gas in the BCS-BEC crossover region is dominated by strong pairing fluctuations, the study of the spin susceptibility in this system would be helpful to see to what extent the preformed pair scenario can explain both the pseudogap phenomenon and the spin-gap phenomenon in a unified manner.
\par
In considering the spin-gap phenomenon in an ultracold Fermi gas, one should note that the ordinary strong-coupling theory developed by Nozi\`eres and Schmitt-Rink\cite{Nozieres,SadeMelo}, as well as the (non-self-consistent) $T$-matrix approximation\cite{Perali2}, that have been extensively used to clarify various BCS-BEC crossover physics in this system, unphysically give negative spin susceptibility in the crossover region\cite{Liu,Parish,Kashimura}. This serious problem has been, however, recently overcome by including higher order pairing fluctuations beyond the $T$-matrix level\cite{Kashimura}. The calculated spin susceptibility in this extended $T$-matrix approximation (ETMA)\cite{Kashimura} agrees well with the recent experiment on a $^6$Li Fermi gas\cite{Sanner}, as well as the theoretical result in the self-consistent $T$-matrix approximation\cite{Enss}. Although the ETMA result disagrees with the experiment done by Sommer and co-workers\cite{Sommer}, it has been pointed out that this experiment corresponds to the case with a repulsive interaction\cite{Taylor,Palestini}. 
\par
In this paper, we employ the ETMA, to calculate the temperature dependence of $\chi_{\rm s}$ over the entire BCS-BEC crossover region. Introducing the spin-gap temperature $T_{\rm s}$ as the temperature below which $\chi_{\rm s}$ is anomalously suppressed, we determine the spin-gap regime in the phase diagram of an ultracold Fermi gas in terms of the temperature and the strength of a pairing interaction. We also deal with the single-particle density of states $\rho(\omega)$ within the same theoretical framework, to clarify how the spin-gap phenomenon is related to the pseudogap phenomenon. 
\par
This paper is organized as follows. In Sec. II, we explain our formulation. Although the ETMA has been explained in Ref. \cite{Kashimura}, we present the outline of this strong-coupling formalism so that our paper can be self-contained. In Sec. III. we show our results on the spin susceptibility, to determine the spin-gap temperature $T_{\rm s}$ in the BCS-BEC crossover region. We also discuss how the spin-gap phenomenon is related to the pseudogap phenomenon there. In Sec. IV, we separately consider the spin-gap phenomenon in the BEC region. Throughout this paper, we set $\hbar=k_{\rm B}=1$, and the system volume $V$ is taken to be unity, for simplicity.  
\par
\section{Formulation}
\par
We consider a uniform two-component Fermi gas, described by the BCS Hamiltonian,
\begin{eqnarray}
\label{eq1}
H=\sum_{\bm{p},\sigma}(\varepsilon_{\bm{p}}-\mu_{\sigma})c^{\dag}_{\bm{p},\sigma}c_{\bm{p},\sigma}
-U\sum_{\bm{p},\bm{p}',\bm{q}}
c^{\dagger}_{\bm{p}+\bm{q}/2,\uparrow}
c^{\dagger}_{-\bm{p}+\bm{q}/2,\downarrow}
c_{-\bm{p}'+\bm{q}/2,\downarrow}
c_{\bm{p}'+\bm{q}/2,\uparrow}.
\end{eqnarray}
Here, $c_{\bm{p},\sigma}$ is the annihilation operator of a Fermi atom with the kinetic energy $\varepsilon_{\bm p}=p^2/2m$ (where $m$ is an atomic mass) and pseudospin $\sigma=\uparrow,\downarrow$ describing two atomic hyperfine states. $\mu_\sigma$ is the Fermi chemical potential in the $\sigma$-component. Although we mainly deal with an unpolarized Fermi gas in this paper, the spin-dependent chemical potential is necessary in calculating the spin susceptibility. $-U$ is an assumed tunable pairing interaction. As usual, we measure the interaction strength in terms of the $s$-wave scattering length $a_{\rm s}$\cite{SadeMelo}, which is related to the bare interaction $-U$ as
\begin{equation}
\label{eq2}
\frac{m}{4\pi a_{\rm s}}=-\frac{1}{U}+\sum_{\bm{p}}^{\omega_{\rm c}}\frac{1}{2\epsilon_{\bm{p}}},
\end{equation}
where $\omega_{\rm c}$ is a cutoff energy. 
\par
\begin{figure}[t]
\begin{center}
\includegraphics[width=6cm]{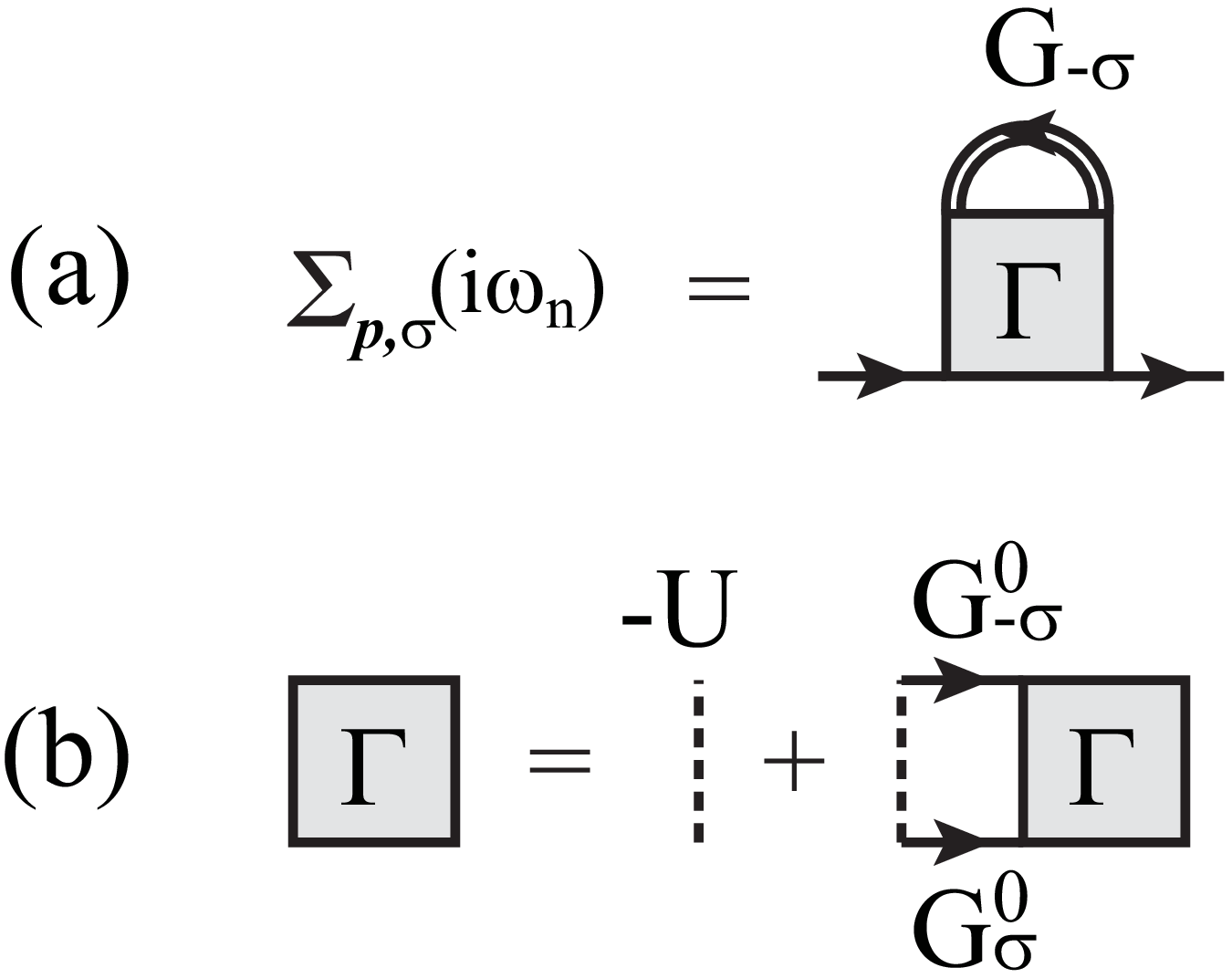}
\end{center}
\caption{(a) Self-energy $\Sigma_{\bm{p},\sigma}(i\omega_{n})$ in the extended $T$-matrix approximation (ETMA). The double solid line and the single solid line are the dressed Green's function $G_{\bm{p},\sigma}(i\omega_{n})$ in Eq. (\ref{eq.3}) and the bare Green's function $G^0_{\bm{p},\sigma}(i\omega_{n})$ in Eq. (\ref{eq.4}), respectively. (b) Particle-Particle scattering matrix $\Gamma_{\bm{q}}(i\nu_{n})$. The dashed line represents the pairing interaction $-U$.}
\label{fig1}
\end{figure}
\par
The extended $T$-matrix approximation (ETMA) is characterized by the self-energy $\Sigma_{\bm{p},\sigma}(i\omega_{n})$ which is diagrammatically given in Fig. \ref{fig1}\cite{Kashimura}. In this figure, the double solid line describes the dressed single-particle thermal Green's function, given by
\begin{equation}
G_{\bm{p},\sigma}(i\omega_{n})=
{1 
\over 
[G^{0}_{\bm{p},\sigma}(i\omega_{n})]^{-1}-\Sigma_{\bm{p},\sigma}(i\omega_{n})
},
\label{eq.3}
\end{equation}
where 
\begin{equation}
G^{0}_{\bm{p},\sigma}(i\omega_{n})=
{1 \over i\omega_{n}-\varepsilon_{\bm{p}}+\mu_{\sigma}}
\label{eq.4}
\end{equation}
is the Green's function for a free Fermi gas (which is described as the single solid line in Fig. \ref{fig1}). In Eqs. (\ref{eq.3}) and  (\ref{eq.4}), $\omega_n$ is the fermion Matsubara frequency. Summing up the ETMA diagrams in Fig. \ref{fig1}, we obtain 
\begin{equation}
\Sigma_{\bm{p},\sigma}(i\omega_{n})=T\sum_{\bm{q},i\nu_{n}}\Gamma_{\bm{q}}(i\nu_{n})G_{\bm{q}-\bm{p},-\sigma}(i\nu_{n}-i\omega_{n}).
\label{eq.5}
\end{equation}
Here, $\nu_{n}$ is the boson Matsubara frequency. The particle-particle scattering matrix $\Gamma_{\bm{q}}(i\nu_{n})$ in Eq. (\ref{eq.5}) has the form,
\begin{equation}
\Gamma_{\bm{q}}(i\nu_{n})=\frac{-U}{1-U\Pi_{\bm{q}}(i\nu_{n})},
\label{eq.6}
\end{equation}
where 
\begin{eqnarray}
\Pi_{\bm{q}}(i\nu_{n})&=&T\sum_{\bm{p},i\nu_{n}}G^{0}_{\bm{p}+\bm{q}/2,\up}(i\nu_{n}+i\omega_{n})G^{0}_{-\bm{p}+\bm{q}/2,\downarrow}(-i\omega_{n})  \nonumber \\
&=&\sum_{\bm{p}}\frac{1-f(\epsilon_{\bm{p}+\bm{q}/2}-\mu_{\up})-f(\epsilon_{-\bm{p}+\bm{q}/2}-\mu_{\downarrow})}{\epsilon_{\bm{p}+\bm{q}/2}+\epsilon_{-\bm{p}+\bm{q}/2}-\mu_{\up}-\mu_{\downarrow}-i\nu_{n}}
\label{eq.7}
\end{eqnarray}
is the lowest order pair correlation function, describing fluctuations in the Cooper channel.
\par
\begin{figure}[t]
\begin{center}
\includegraphics[width=6cm]{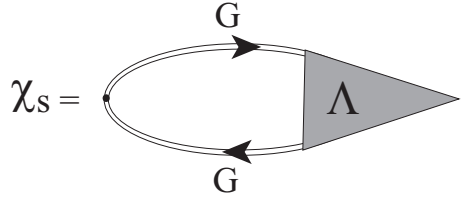}
\end{center}
\caption{Feynman diagram describing spin susceptibility $\chi_{\rm s}$. The dressed Green's function $G$ involves the self-energy correction. $\Lambda$ is a three-point spin-vertex function. The filled circle represents the bare spin-vertex part.}
\label{fig2}
\end{figure}
\par
Within the framework of the ETMA, the uniform spin susceptibility $\chi_{\rm s}$ can be conveniently calculated from
\begin{equation}
\chi = \lim_{h\rightarrow 0}\frac{\Delta N}{h}\equiv\lim_{h\rightarrow 0}\frac{N_{\up}-N_{\downarrow}}{h},
\label{eq.8}
\end{equation}
where $h=\mu_\uparrow-\mu_\downarrow$ is a fictitious magnetic field. The number $N_\sigma$ of Fermi atoms in the $\sigma$-component is given by
\begin{equation}
N_{\sigma}=T\sum_{\bm{p},i\omega_{n}}G_{\bm{p},\sigma}(i\omega_{n}).
\label{eq.9}
\end{equation}
The advantage of using Eq. (\ref{eq.8}) is that the vertex correction $\Lambda$ in Fig. \ref{fig2} which is consistent with the ETMA self-energy in Eq. (\ref{eq.5}) is automatically taken into account. Thus, the Ward identity is satisfied, which is a required condition for any consistent theory. In this paper, we numerically evaluate Eq. (\ref{eq.9}) by taking a small but finite value, $h/\varepsilon_{\rm F}=O(10^{-2})$\cite{note} (where $\varepsilon_{\rm F}$ is the Fermi energy). 
\par
In this paper, we also consider the single-particle density of states $\rho(\omega)$, to examine the relation between the spin-gap phenomenon and the pseudogap phenomenon. This quantity is calculated from the analytic continued Green's function as
\begin{equation}
\rho(\omega)=-\frac{1}{\pi}\sum_{\bm{p}}{\rm Im}[G_{\bm p}(i\omega_{n}\rightarrow\omega+i\delta)],
\label{eq.10}
\end{equation}
where $\delta$ is an infinitesimally small positive number. Since we are considering an unpolarized Fermi gas ($\mu_\uparrow=\mu_\downarrow\equiv\mu$), we have suppressed the spin index $\sigma$ in Eq. (\ref{eq.10}). We always use this simplified notation in this paper, when we deal with the unpolarized case.
\par
The superfluid instability is determined from the Thouless criterion, $[\Gamma_{\bm{q}=0}(i\nu_{n}=0)]^{-1}=0$\cite{Thouless}, which gives
\begin{equation}
1={U \over 2}\sum_{\bm p}
{\displaystyle \tanh{\varepsilon_{\bm{p}}-\mu \over 2T_{\rm c}} \over \varepsilon_{\bm p}-\mu}.
\label{eq.10b}
\end{equation}
We numerically solve the $T_{\rm c}$-equation (\ref{eq.10b}), together with the number equation (\ref{eq.9}), under the condition $N_\uparrow=N_\downarrow=N/2$ (where $N$ is the total number of Fermi atoms). We self-consistently determine $T_{\rm c}$ and $\mu$ for a given interaction strength.
\par
Before ending section, we briefly note that the self-energy in the ordinary (non-self-consistent) $T$-matrix approximation (TMA) is given by replacing the dressed Green's function $G$ in Eq. (\ref{eq.5}) with the bare one $G^0$. Because of this improvement, the ETMA correctly gives positive $\chi_{\rm s}$ in the whole BCS-BEC crossover region\cite{Kashimura}. In contrast, the TMA spin susceptibility unphysically becomes negative in the interesting BCS-BEC crossover region. For more details of this difference, we refer to Ref. \cite{Kashimura}.
\par
\par
\section{Spin-gap phenomenon and relation to pseudogap phenomenon in an ultracold Fermi gas}
\par
\begin{figure}[t]
\begin{center}
\includegraphics[width=8cm]{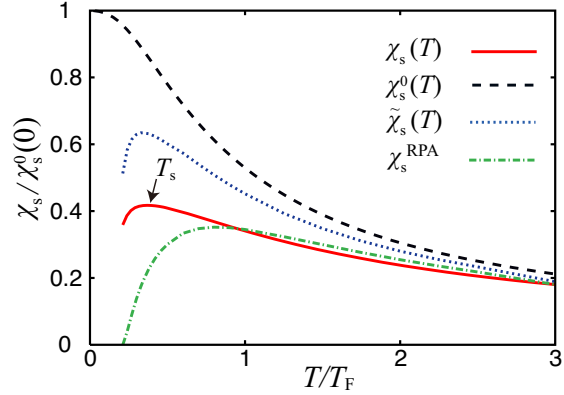}
\end{center}
\caption{(Color online) Calculated ETMA uniform spin susceptibility $\chi_{\rm s}(T)$ in a unitary Fermi gas ($T_{\rm c}=0.21T_{\rm F}$), as a function of temperature. $\chi_{\rm s}^0(T)$ is the spin susceptibility in the case of a free Fermi gas and $T_{\rm F}$ is the Fermi temperature. ${\tilde \chi}_{\rm s}(T)$ is given in the second line in Eq. (\ref{eq.13}). $\chi_{\rm s}^{\rm RPA}(T)$ is given in Eq. (\ref{eq.11b}). The arrow shows the spin-gap temperature $T_{\rm s}$.}
\label{fig3}
\end{figure}

Figure \ref{fig3} shows the uniform spin susceptibility $\chi_{\rm s}(T)$ in a unitary Fermi gas above $T_{\rm c}$ ($(k_{\rm F}a_{\rm s})^{-1}=0$, where $k_{\rm F}$ is the Fermi momentum). In this figure, one sees that $\chi_{\rm s}(T)$ exhibits  non-monotonic temperature dependence. While the temperature dependence of $\chi_{\rm s}(T\ge 0.37T_{\rm F})$ is qualitatively the same as the non-interacting case ($\chi_{\rm s}^0(T)$), $\chi_{\rm s}(T)$ anomalously decreases with decreasing the temperature when $T_{\rm c}\le T\le 0.37T_{\rm F}$. Since the latter phenomenon never occurs in the non-interacting case, it is considered to originate from strong pairing fluctuations near $T_{\rm c}$. We briefly note that this low temperature behavior of $\chi_{\rm s}$ is analogous to the spin-gap phenomenon observed in the under-doped regime of high-$T_{\rm c}$ cuprates. We also note that the non-monotonic temperature dependence of the spin susceptibility in a unitarity Fermi gas has been also obtained by the self-consistent $T$-matrix approximation\cite{Enss}, as well as quantum Monte-Carlo simulation\cite{Bulgac}.
\par

\begin{figure}[t]
\begin{center}
\includegraphics[width=8cm]{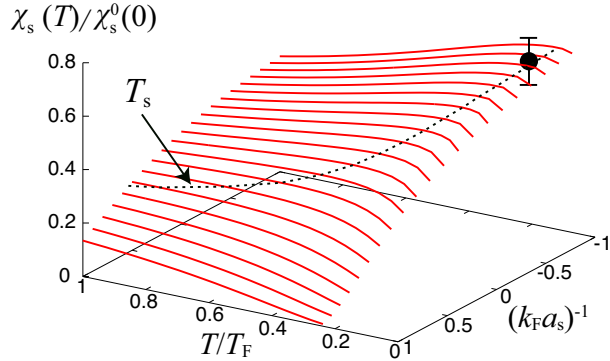}
\end{center}
\caption{(Color online) Calculated spin susceptibility $\chi_{\rm s}$ in the BCS-BEC crossover region above $T_{\rm c}$ (solid line). The dashed line shows the peak position of $\chi_{\rm s}$, which gives the spin-gap temperature $T_{\rm s}$. The filled circle with error bar is the observed spin susceptibility in a $^6$Li Fermi gas at $(k_{\rm F}a_{\rm s})^{-1}\simeq -0.8$\cite{Sanner} (where $k_{\rm F}$ is the Fermi momentum).}
\label{fig4}
\end{figure}

\begin{figure}[t]
\begin{center}
\includegraphics[width=8cm]{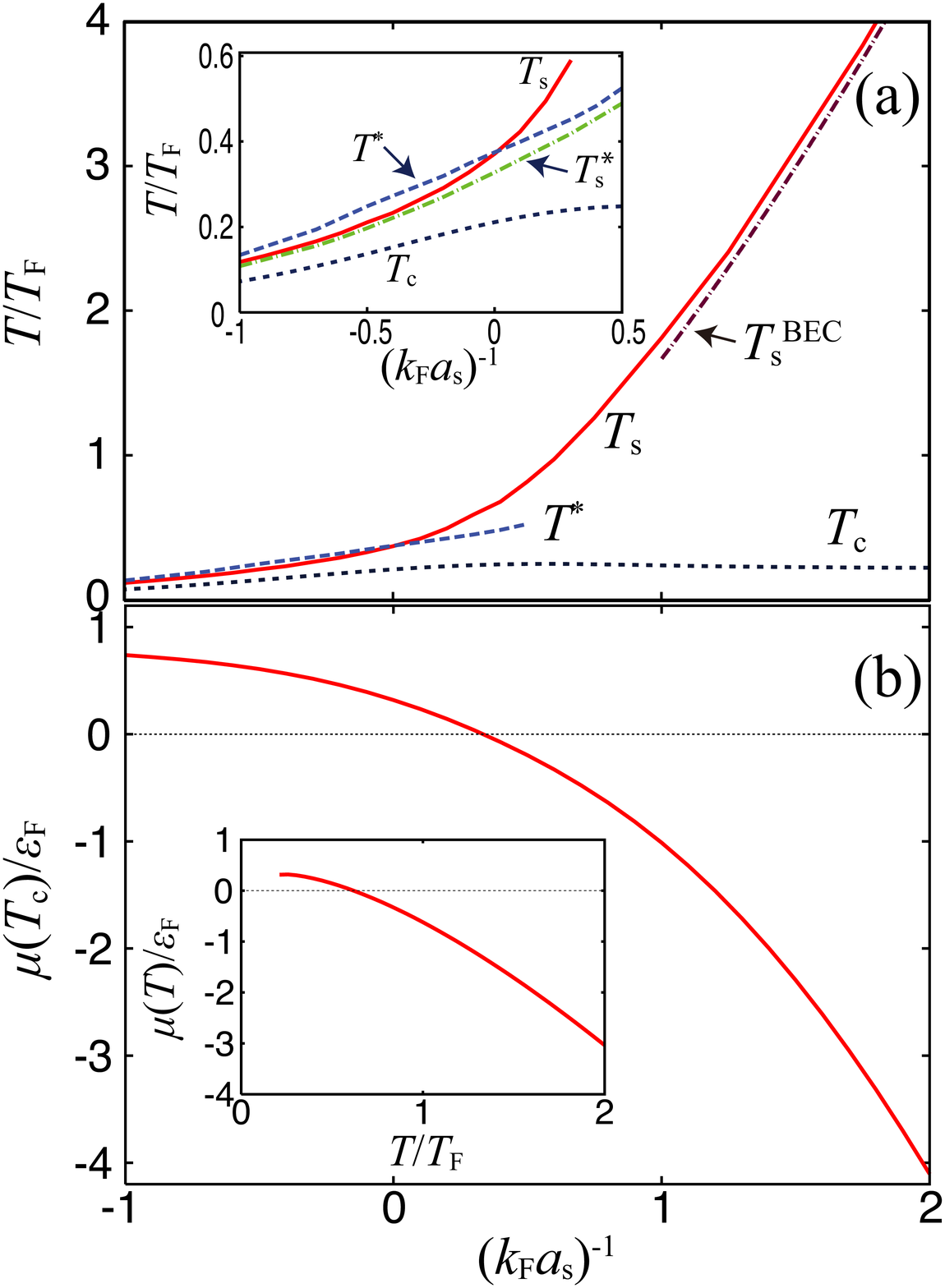}
\end{center}
\caption{(Color online) (a) Spin-gap temperature $T_{\rm s}$ in the BCS-BEC crossover regime of an ultracold Fermi gas. $T^*$ is the pseudogap temperature which is defined as the temperature at which the density of states $\rho(\omega=0)$ takes a maximum value. $T_{\rm s}^{\rm BEC}$ is obtained from Eq. (\ref{eq.17}). The inset shows the result magnified in the BCS side, where $T_{\rm s}^*$ is the spin-gap temperature determined from the approximate spin susceptibility ${\tilde \chi}_{\rm s}$ given by the second line in Eq. (\ref{eq.13}). (b) Calculated Fermi chemical potential $\mu$ at $T_{\rm c}$. The inset shows the temperature dependence of $\mu$ in the unitarity limit.}
\label{fig5}
\end{figure}
\par
The non-monotonic temperature dependence of $\chi_{\rm s}$ is obtained over the entire BCS-BEC crossover region, as shown in Fig. \ref{fig4}. Using this, we conveniently introduce the spin-gap temperature $T_{\rm s}$ as the temperature at which $\chi_{\rm s}$ takes a maximum value.  As shown in Fig. \ref{fig5}(a), the spin-gap temperature $T_{\rm s}$ monotonically increases with increasing the interaction strength. Although $T_{\rm s}$ is a characteristic temperature without being accompanied by any phase transition, we simply call the temperature region $T_{\rm c}\le T\le T_{\rm s}$ the spin-gap regime, where the spin excitations are suppressed.
\par
In Fig. \ref{fig4}, we also plot the experimental result on a $^6$Li Fermi gas at $(k_{\rm F}a_{\rm s})^{-1}\simeq -0.8$ (filled circle)\cite{Sanner}. Besides the good agreement of our result with the observed spin susceptibility\cite{Kashimura}, we find that the experimental result is nearly located at the spin gap temperature $T_{\rm s}$. Thus, when one varies the temperature in this experiment, the decrease of the spin susceptibility is expected, which would be a useful experimental check to confirm the existence of the spin-gap phenomenon.
\par
\begin{figure}[t]
\begin{center}
\includegraphics[width=7cm,height=14cm]{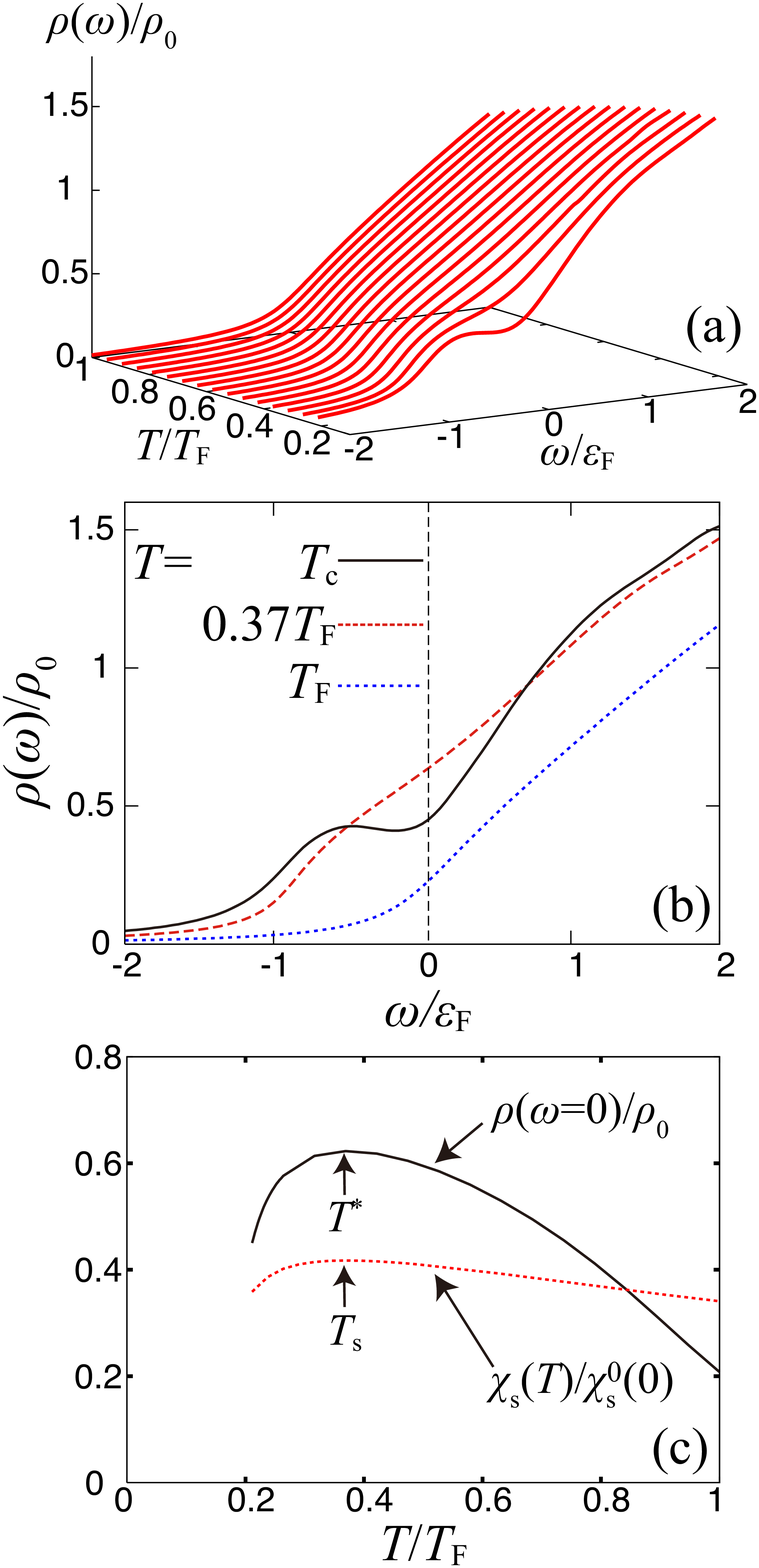}
\end{center}
\caption{(Color online) (a) Calculated density of states $\rho(\omega)$ in the unitarity limit. In this panel, the rightmost line shows the result at $T_{\rm c}$, where one sees a pseudogap (dip) structure around $\omega=0$. $\rho_0=mk_{\rm F}/(2\pi^2)$ is the density of states at the Fermi level in a free Fermi gas. (b) Density of states $\rho(\omega)$ at some typical temperatures in the unitarity limit. $\rho(\omega=0)$ takes a maximum value at $T=0.37T_{\rm F}$, which we call the pseudogap temperature $T^*$ in this paper. (c) Evaluated $\rho(\omega=0)$ as a function of temperature in the unitarity limit. For comparison, we also plot $\chi_{\rm s}$ in this panel, where one sees $T_{\rm s}\simeq T^*$.}
\label{fig6}
\end{figure}
\par
As shown in Ref. \cite{Kashimura}, the ETMA also gives pseudogapped density of states $\rho(\omega)$ in the unitarity limit. In Fig. \ref{fig6}(a), one sees that a dip structure in $\rho(\omega\simeq 0)$ gradually disappears with increasing the temperature from $T_{\rm c}$. To see how this pseudogap phenomenon is related to the spin-gap phenomenon in a simply manner, it is helpful to approximately evaluate the spin susceptibility within the neglect of the spin-vertex correction $\Lambda$ in Fig. \ref{fig2}, which gives ($\equiv {\tilde \chi}_{\rm s}$)
\begin{eqnarray}
{\tilde \chi}_{\rm s}
=-T\sum_{{\bm p},\omega_n}G^2_{\bm p}(i\omega_n)
=-\sum_{\bm p}\int_{-\infty}^\infty dz \int_{-\infty}^\infty dz' 
      A_{\bm p}(z)A_{\bm p}(z')\frac{f(z)-f(z')}{z -z'}.
\label{eq.11}
\end{eqnarray}
Here, $f(z)=[\exp({z/T})+1]^{-1}$ is the Fermi distribution function, and $A_{\bm p}(z)=-{\rm Im}[G_{\bm p}(i\omega_n\to z+i\delta)]/\pi$ is the single-particle spectral weight. When the quasiparticle damping described by the imaginary part of the analytic continued self-energy ${\rm Im}[\Sigma({\bm p},i\omega_n\to z+i\delta$)] is weak (which is justified in the weak-coupling regime), the factor $A_{\bm p}(z)A_{\bm p}(z')$ in Eq. (\ref{eq.11}) becomes large only when $z\simeq z'$. In this case, Eq. (\ref{eq.11}) is reduced to
\begin{eqnarray}
{\tilde \chi}_{\rm s}
&\simeq&
-\sum_{\bm p}\int_{-\infty}^\infty dz A_{\bm p}(z){df(z) \over dz}
\int_{-\infty}^\infty dz' A_{\bm p}(z')
\nonumber
\\
&=&\int_{-\infty}^\infty dz \rho(z)
\left(-{df(z) \over dz}\right)
\nonumber
\\
&\simeq&
\rho(0),
\label{eq.13}
\end{eqnarray}
where $\rho(z)=\sum_{\bm p}A_{\bm p}(z)$ is just the density of states in Eq. (\ref{eq.10}). In obtaining the last expression in Eq. (\ref{eq.13}), we have employed the approximation, 
\begin{equation}
-{df(z) \over dz}\simeq\delta(z).
\label{eq.FFF}
\end{equation} 
\par
Equation (\ref{eq.13}) indicates that the spin-gap behavior of the spin susceptibility directly reflects the temperature dependence of the pseudogapped density of states around $\omega=0$. Indeed, Fig. \ref{fig6}(b) indicates that $\rho(\omega=0)$ exhibits non-monotonic temperature dependence. When we introduce the characteristic temperature $T^*$ at which $\rho(\omega=0)$ takes a maximum value, Fig. \ref{fig6}(c) shows that $T_{\rm s}\simeq T^*$, as expected. Since the suppression of $\rho(\omega=0)$ below $T^*$ is due to the formation of the pseudogap, the spin-gap seen in Fig. \ref{fig3} is found to originate from the pseudogap phenomenon.
\par
We note that the present definition of the pseudogap temperature $T^*$ is a bit different from the ordinary one that a dip structure appears in $\rho(\omega\simeq 0)$ below $T^*$\cite{Tsuchiya,Watanabe}. Since the pseudogap is a crossover phenomenon without being accompanied by any phase transition, the definition of the pseudogap temperature always involves ambiguity to some extent. In this regard, we point out that the coincidence of $T_{\rm s}$ and $T^*$ shown in Fig. \ref{fig6}(c) indicates that the present definition is convenient in considering the relation between the spin-gap phenomenon and the pseudogap phenomenon. 
\par
We also note that, while the low temperature behavior of $\rho(\omega=0,T\le T^*)$ in Fig. \ref{fig4} is due to the pseudogap phenomenon, the high temperature behavior simply comes from the temperature dependence of the chemical potential $\mu$, which has been already seen in a free Fermi gas. In the non-interacting case, the Fermi chemical potential far below $T_{\rm F}$ is given by\cite{Kubo}
\begin{equation}
\mu(T)\simeq \varepsilon_{\rm F}
\left[
1-{\pi^2 \over 12}
\left(
{T \over T_{\rm F}}
\right)^2
\right].
\label{eq.14}
\end{equation}
(A similar temperature dependence of $\mu$ is also obtained in the unitarity limit, as shown in the inset in Fig. \ref{fig5}(b).) Thus, the density of states $\rho_0(\omega=0)$ in a free Fermi gas decreases with increasing the temperature as
\begin{equation}
\rho_{0}(\omega=0)=
{m \over 2\pi^2}\sqrt{2m\mu(T)}
\simeq
{mk_{\rm F} \over 2\pi^2}
\sqrt{
1-{\pi^2 \over 12}
\left(
{T \over T_{\rm F}}
\right)^2}.
\label{eq.15}
\end{equation}
\par
\begin{figure}[t]
\begin{center}
\includegraphics[width=8cm]{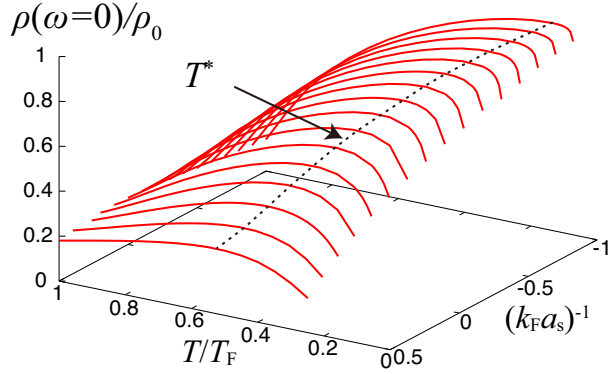}
\end{center}
\caption{Calculated density of states $\rho(\omega=0)$ in the BCS-BEC crossover region above $T_{\rm c}$. The dotted line shows the pseudogap temperature $T^*$ at which $\rho(0)$ takes a maximum value.}
\label{fig7}
\end{figure}
\par
Figure \ref{fig7} shows the density of states $\rho(\omega=0)$ in the BCS-BEC crossover region. Determining $T^*$ from this figure, one finds in Fig. \ref{fig5}(a) that $T^*$ agrees well with the spin-gap temperature $T_{\rm s}$, not only in the unitarity limit, but also in the BCS side $(k_{\rm F}a_{\rm s})^{-1}\le 0$. Although $T^*$ is slightly higher than $T_{\rm s}$ in the BCS regime (See the inset in Fig. \ref{fig5}(a).), it is simply due to the approximation in Eq. (\ref{eq.FFF}) in obtaining the last expression in Eq. (\ref{eq.13}). When we substitute the ETMA density of states into the second line in Eq. (\ref{eq.13}), and evaluate the spin-gap temperature ($\equiv T_{\rm s}^*$), $T_{\rm s}^*$ well reproduces $T_{\rm s}$ in the BCS side, as shown in the inset in Fig. \ref{fig5}(a). Since there is no experimental technique to directly measure $\rho(\omega)$ in cold Fermi gas physics, our result indicates that the observation of the spin-gap temperature $T_{\rm s}$ is a useful approach to detect the presence of the pseudogap, as least in the BCS side. 
\par
\begin{figure}[t]
\begin{center}
\includegraphics[width=8cm]{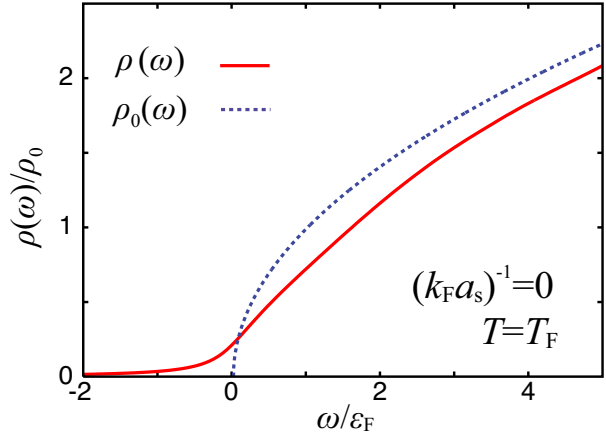}
\end{center}
\caption{(Color online) Calculated ETMA density of states $\rho(\omega)$ at $T_{\rm F}$ in a unitary Fermi gas (solid line). The dashed line shows the density of states $\rho_0(\omega)=(m\sqrt{2m}/2\pi^2)\sqrt{\omega+\mu(T=T_{\rm F})}$ in a free Fermi gas.}
\label{fig8}
\end{figure}
\par
Before ending this section, we briefly explain the reason why $\chi_{\rm s}(T)$  in Fig. \ref{fig3} is smaller than the non-interacting result $\chi_{\rm s}^0(T)$ even far above $T_{\rm c}$. When $T\gesim T_{\rm F}$, although pairing fluctuations are weak, particle-particle scatterings still modify the single-particle excitation spectrum, leading to the modification of the density of states $\rho(\omega)$. Indeed, Fig. \ref{fig8} shows that $\rho(\omega,T=T_{\rm F})$ in the unitarity limit is different from the density of states $\rho_0(\omega)$ in the case of a free Fermi gas at the same temperature. Using this modified density of states $\rho(\omega)$ in evaluating the second line in Eq. (\ref{eq.13}), one obtains ${\tilde \chi}_{\rm s}<\chi_{\rm s}^0$. (See Fig. \ref{fig3}.) The reason why ${\tilde \chi}_{\rm s}$ is still larger than the ETMA spin susceptibility $\chi_{\rm s}$ is simply due to the vertex correction $\Lambda$ ignored in Eq. (\ref{eq.13}). As pointed out in Ref. \cite{Kashimura}, when we simply approximate the particle-particle scattering matrix $\Gamma_{\bm q}(i\nu)$ to the value in the low-energy and low-momentum limit ($\equiv U_{\rm eff}$), the ETMA vertex correction $\Lambda$ involves the RPA (random phase approximation)-type Stoner factor\cite{Yosida}. Extending Eq. (\ref{eq.13}) to include this vertex correction, we obtain
\begin{equation}
\chi_{\rm s}^{\rm RPA}(T)=
{{\tilde \chi}_{\rm s}(T) \over 1-U_{\rm eff}{\tilde \chi}_{\rm s}(T)}.
\label{eq.11b}
\end{equation}
Because $U_{\rm eff}<0$ in the present attractive case ($-U<0$), the Stoner factor suppresses the spin susceptibility. As shown in Fig. \ref{fig3}, Eq. (\ref{eq.11b}) well describes $\chi_{\rm s}$ in the high temperature region. Although such agreement is not obtained when $T\lesssim T_{\rm F}$, this is because vertex corrections beyond the RPA become crucial there, due to pairing fluctuations enhanced near $T_{\rm c}$. 
\par
\par
\section{Spin-gap phenomenon in the BEC regime}
\par
When $(k_{\rm F}a_{\rm s})^{-1}\gesim 0$, $T_{\rm s}$ gradually deviates from the pseudogap temperature $T^*$, as seen in Fig. \ref{fig5}(a). In this strong-coupling regime, one cannot ignore the quasiparticle damping effect (which is described by the imaginary part of the analytic continued self-energy), as well as the vertex correction $\Lambda$ in Fig. \ref{fig2}, so that Eq. (\ref{eq.13}) is no longer valid. 
\par
To understand physics behind $T_{\rm s}$ in the BEC regime, we recall that this regime may be viewed as a molecular Bose gas. Since these bound molecules are in the spin-singlet state, they do not contribute to the spin susceptibility. Thus, $\chi_{\rm s}$ in this regime is dominated by Fermi atoms associated with thermal dissociation of molecules. When one approximately treats this situation as a gas mixture of $N_{\rm M}$ free spinless Bose molecules and $N_\sigma^{\rm F}$ free Fermi atoms ($\sigma=\uparrow,\downarrow$), $\chi_{\rm s}$ is evaluated as
\begin{equation} 
\chi_{\rm s}= \lim_{h\rightarrow 0}\frac{N^{\rm F}_{\up}-N^{\rm F}_{\downarrow}}{h}.
\label{eq.16}
\end{equation}
Using this, one obtains the equation for the spin-gap temperature ($\equiv T_{\rm s}^{\rm BEC}$) as
\begin{equation}
\frac{1}{\sqrt{2}}\frac{(2\pi m T_{\rm s}^{\rm BEC})^{\frac{3}{2}}}{(2\pi)^{3}N}\exp\left(-\frac{E_{\rm b}}{T_{\rm s}^{\rm BEC}}\right)=4\left[\left(\frac{2E_{\rm b}+3T_{\rm s}^{\rm BEC}}{2E_{\rm b}-T_{\rm s}^{\rm BEC}}\right)^{2}-1\right]^{-1},
\label{eq.17}
\end{equation}
where $E_{\rm b}=1/(ma_{\rm s}^{2})$ is the binding energy of a two-body bound state\cite{SadeMelo}. (We summarize the derivation of Eq. (\ref{eq.17}) in the Appendix.) As shown in Fig. \ref{fig5}(a), $T_{\rm s}^{\rm BEC}$ well describes the spin-gap temperature $T_{\rm s}$ in the BEC regime. This means that $T_{\rm s}$ in this regime is dominated by the thermal dissociation of two-body bound molecules. 
\par
We point out that Eq. (\ref{eq.17}) is very similar to the famous Saha's equation in classical plasma physics\cite{Saha},
\begin{equation}
\frac{1}{\sqrt{2}}\frac{(2\pi m T_{\rm Saha})^{\frac{3}{2}}}{(2\pi)^{3}N}\exp\left(-\frac{E_{\rm I}}{T_{\rm Saha}}\right)=\frac{\alpha^{2}}{1-\alpha},
\label{eq.18}
\end{equation}
where $E_{\rm I}$ is the ionization energy of a particle, which corresponds to the binding energy $E_{\rm b}$ in Eq. (\ref{eq.17}). The Saha's equation (\ref{eq.18}) determines the dissociation temperature $T_{\rm Saha}$ at which a given value of the ionization rate $\alpha$ is achieved. The fact of $T_{\rm s}\simeq T_{\rm s}^{\rm BEC}$, as well as the similarity between Eqs. (\ref{eq.17}) and (\ref{eq.18}), indicate that the spin-gap temperature $T_{\rm s}$ in the BEC regime is physically similar to the Saha's temperature $T_{\rm Saha}$ discussed in classical plasma physics. 
\par
\begin{figure}[t]
\begin{center}
\includegraphics[width=8cm]{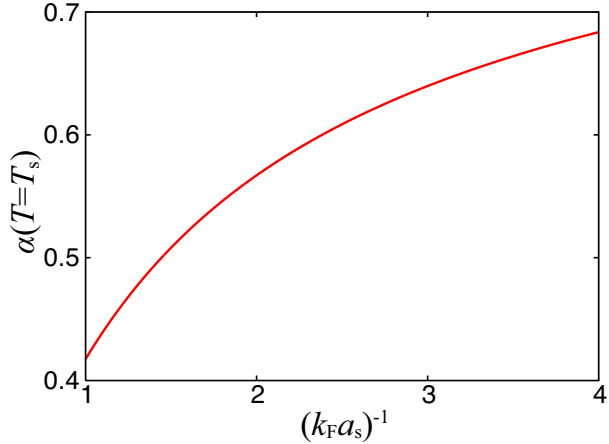}
\end{center}
\caption{(Color online) Calculated dissociation rate $\alpha(T=T_{\rm s})$ in the BEC regime obtained by equating the right hand sides of Eqs. (\ref{eq.17}) and (\ref{eq.18}).}
\label{fig9}
\end{figure}
\par
Because of the similarity between Eqs. (\ref{eq.17}) and (\ref{eq.18}), it is interesting to evaluate the ``dissociation rate" $\alpha=[N_\uparrow^{\rm F}+N_\downarrow^{\rm F}]/N$ in the present case. Equating the right hand sides of these equations, we find that $\alpha(T_{\rm s})\simeq 0.5$ when $(k_{\rm F}a_{\rm s})^{-1}\sim 1.5$, as shown in Fig. \ref{fig9}. This means that, around this interaction strength, bound molecules are thermally dissociated into Fermi atoms at $T\sim T_{\rm s}$. We briefly note that $\alpha(T_{\rm s})$ approaches unity in the extreme BEC limit ($(k_{\rm F}a_{\rm s})^{-1}\to\infty$).
\par

\section{Summary}
\par
To summarize, we have discussed spin-gap phenomena in the normal state of an ultracold Fermi gas. Including strong pairing fluctuations within the framework of an extended $T$-matrix approximation, we have calculated the uniform spin susceptibility $\chi_{\rm s}$ in the BCS-BEC crossover region. We showed that $\chi_{\rm s}$ exhibits non-monotonic temperature dependence. In particular, $\chi_{\rm s}$ is anomalously suppressed near $T_{\rm c}$, which is similar to the spin-gap phenomenon known in high-$T_{\rm c}$ cuprates. To characterize the spin-gap phenomenon in the present case, we have introduced the spin-gap temperature $T_{\rm s}$ as the temperature at which $\chi_{\rm s}$ takes a maximum value. Determining $T_{\rm s}$ over the entire BCS-BEC crossover region, we have identified the spin-gap regime, which is wider for a stronger pairing interaction, as expected. 
\par
We clarified how the spin-gap phenomenon is related to the pseudogap phenomenon appearing in the single-particle density of states $\rho(\omega)$. Introducing the pseudogap temperature $T^*$ as the temperature at which $\rho(\omega=0)$ takes a maximum value, we found that $T^*$ agrees well with $T_{\rm s}$, when $(k_{\rm F}a_{\rm s})^{-1}\lesssim 0$. Since the suppression of $\rho(\omega\simeq 0)$ below $T^*$ is characteristic of the pseudogap phenomenon, this agreement means that the spin-gap phenomenon in the BCS side originates from the pseudogapped density of states.
\par
The spin-gap temperature $T_{\rm s}$ gradually deviates from $T^*$, as one enters the BEC side ($(k_{\rm F}a_{\rm s})^{-1}\gesim 0$). In this strong-coupling regime, the system may be viewed as a gas mixture of tightly bound molecules and unpaired Fermi atoms. Since the former molecules are in the spin-singlet state, the latter fermions only contribute to the spin susceptibility. Indeed, we showed that $T_{\rm s}$ in the BEC regime is well described by the spin-gap temperature $T_{\rm s}^{\rm BEC}$ which is evaluated in a model gas mixture of two-component free fermions and free spinless bosons. We also showed that the equation for $T_{\rm s}^{\rm BEC}$ is similar to the Saha's equation in classical plasma physics. The good agreement of $T_{\rm s}$ with $T_{\rm s}^{\rm BEC}$, as well as the similarity between the equation for $T_{\rm s}^{\rm BEC}$ and the Saha's equation, indicate that the spin-gap phenomenon in the BEC regime is dominated by thermal dissociation of tightly bound molecules. Evaluating the dissociation rate $\alpha(T=T_{\rm s})$, we obtain $\alpha(T_{\rm s})\sim 0.5$ when $(k_{\rm F}a_{\rm s})^{-1}\sim 1.5$.
\par
In cold Fermi gas physics, although it is a crucial problem whether the pseudogap really exists or not, there is no experimental technique to directly measure the single-particle density of states, which makes the pseudogap problem difficult. Since the uniform spin susceptibility $\chi_{\rm s}$ is observable in this system, our result would be useful for the detection of the pseudogap phenomenon through this quantity in this system. In addition, the pseudogap phenomenon and spin-gap phenomenon are crucial keys to clarify the pairing mechanism of high-$T_{\rm c}$ cuprates. In this regard, our results would also contribute to the understanding of these many-body phenomena in a unified manner, using ultracold Fermi gases.
\par
\acknowledgements
We would like to thank D. Inotani for useful discussions. Y.O. was supported by Grant-in-Aid for Scientific Research from MEXT in Japan (25400418, 25105511, 23500056).
\par
\appendix
\section{Derivation of Eq. (\ref{eq.17})}
\par
In the classical regime ($T\gg T_{\rm F}$), $N_\sigma^{\rm F}$ and $N_{\rm M}$ are given by, respectively,
\begin{eqnarray}
N^{\rm F}_{\sigma}=
\sum_{\bm{p}}\exp{\left(-\frac{\varepsilon_{\bm{p}}-\mu-\sigma h/2}{T}\right)}
=\frac{3\sqrt{\pi}N}{8}
\left({T \over \varepsilon_{\rm F}}\right)^{3/2}
\lambda\exp{\left(\frac{\sigma h}{2T}\right)},
\label{eq.A1}
\end{eqnarray}
\begin{eqnarray}
N_{\rm M}=\sum_{\bm{q}}\exp{\left(-\frac{\varepsilon_{\bm{q}}/2-2\mu-E_{\rm b}}{T}\right)}
&=&\frac{3\sqrt{2\pi}N}{4}
\left({T \over \varepsilon_{\rm F}}\right)^{3/2}
\lambda^{2}\exp{\left(\frac{E_{\rm b}}{T}\right)}.
\label{eq.A2}
\end{eqnarray}
Here, $E_{\rm b}=1/(ma_{\rm s}^2)$ is the binding energy of a molecule, and $\lambda=\exp{(\mu/T)}$ is the fugacity. Substituting Eqs. (\ref{eq.A1}) and (\ref{eq.A2}) into the number equation, $N=N_{\uparrow}^{\rm F}+N_\downarrow^{\rm F}+2N_{\rm M}$, one obtains
\begin{eqnarray}
\label{eq.A3}
\lambda=\frac{1}{4\sqrt{2}}\exp{\left(-\frac{E_{\rm b}}{T}\right)}
\left[\sqrt{1+\frac{32}{3}\sqrt{\frac{2}{\pi}}
\left({T \over \varepsilon_{\rm F}}\right)^{-3/2}
\exp{\left(\frac{E_{\rm b}}{T}\right)}}-1\right].
\end{eqnarray}
Evaluating Eq. (\ref{eq.16}) using Eq. (\ref{eq.A1}), we obtain
\begin{equation} 
\chi_{\rm s}=\frac{\sqrt{\pi}}{2}\sqrt{T \over \varepsilon_{\rm F}}
\lambda\chi_{\rm s}^0(0),
\label{eq.A4}
\end{equation}
where $\chi_{\rm s}^0(0)$ is the spin susceptibility in a free Fermi gas at $T=0$. The spin-gap temperature $T_{\rm s}^{\rm BEC}$ is determined from the condition $\partial\chi_{\rm s}/\partial T=0$, which gives
\begin{eqnarray}
\sqrt{\varepsilon_{\rm F} \over T}\lambda
+2\sqrt{\varepsilon_{\rm F}T}
{\partial\lambda \over \partial T}
=0.
\label{eq.A5}
\end{eqnarray}
In Eq. (\ref{eq.A5}), the derivative of $\lambda$ in terms of $T$ is given by 
\begin{eqnarray}
\frac{\partial \lambda}{\partial T}=\frac{3\lambda}{2T}
{\displaystyle
\left(\frac{4\sqrt{2}}{3}\frac{E_{\rm b}}{T}-2\sqrt{2}\right)\lambda\exp{\left(\frac{E_{\rm b}}{T}\right)}-1
\over
\displaystyle
4\sqrt{2}\lambda\exp{\left(\frac{E_{\rm b}}{T}\right)}+1}.
\label{eq.A6}
\end{eqnarray}
Substituting Eqs. (\ref{eq.A3}) and (\ref{eq.A6}) into Eq. (\ref{eq.A5}), we obtain Eq. (\ref{eq.17}).
\par
We briefly note that the Saha's equation (\ref{eq.18}) is obtained, when one divides $\alpha^2=[N^{\rm F}_\uparrow+N^{\rm F}_\downarrow]^2/N^2$ by $2N_{\rm M}/N~(= 1-\alpha)$ with $h=0$.
\par


\begin{thebibliography}{9}
\bibitem{note} We have numerically confirmed that the ``magnetization'' $N_\uparrow-N_\downarrow$ is proportional to $h$, when $h/\varepsilon_{\rm F}=O(10^{-2})$.
\end{thebibliography}
\end{document}